\documentstyle[12pt,epsfig]{article}
\newlength{\dinwidth}
\newlength{\dinmargin}
\def\bq{\begin{equation}}
\def\eq{\end{equation}}
\def\bqa{\begin{eqnarray}}
\def\eqa{\end{eqnarray}}
\def\bqb{\begin{eqnarray*}}
\def\eqb{\end{eqnarray*}}
\def\to{\rightarrow}

\def\slepton{\widetilde \ell}

\def\slr{\slepton_{R}}
\def\sll{\slepton_{L}}
\def\squark{\widetilde q}

\def\sneu{\widetilde \nu}
\def\sele{\widetilde e}

\def\ser{\sele_{R}}

\def\sql{\squark_{L}}

\def\msn{m_{\sneu}}

\def\sd{\widetilde{d}}

\def\sdr{\sd_{R}}

\def\sz{{\widetilde{\chi}}^0}
\def\sz1{{\widetilde{\chi}}^0_{1}}

\def\mser{m_{\ser}}
\def\msz1{m_{\sz1}}

\def\nle{\mathrel{\vcenter
     {\hbox{$<$}\nointerlineskip\hbox{$\sim$}}}}
\def\nge{\mathrel{\vcenter
     {\hbox{$>$}\nointerlineskip\hbox{$\sim$}}}}

\def\rb{\slash\hspace{-7pt}R}

\begin{document}
~~~\\
\vspace{10mm}
\begin{flushright}
AP-SU-01/02  \\
OCHA-PP-187 \\
hep-ph/0200000 \\
Mar. 2002
\end{flushright}
\begin{center}
  \begin{Large}
   \begin{bf}
Scalar neutrino production in R-parity violating supersymmetric model at HERA\\
   \end{bf}
  \end{Large}
  \vspace{5mm}
  \begin{large}
Tadashi Kon \\
  \end{large}
Faculty of Engineering, Seikei University, Tokyo 180-8633, Japan \\
kon@apm.seikei.ac.jp\\
 \vspace{3mm}
  \begin{large}
Jun-ichi Kamoshita \\
  \end{large}
Department of Physics, Ochanomizu University, Tokyo 112-8610, Japan \\
kamosita@sofia.phys.ocha.ac.jp\\
 \vspace{3mm} \begin{large}
    Tetsuro Kobayashi\\
  \end{large}
Faculty of Engineering, 
Fukui Institute of Technology, Fukui 910-8505, Japan \\
koba@ge.seikei.ac.jp\\
 \vspace{3mm}
 \begin{large}
    Shoichi Kitamura\\
  \end{large}
Tokyo Metropolitan University of Health Sciences, Tokyo 116-8551, Japan\\
    kitamura@post.metro-hs.ac.jp\\
 \vspace{3mm}
  \begin{large}
    Yoshimasa Kurihara\\
  \end{large}
KEK, Tsukuba 305-0801, Japan\\
    kurihara.yoshimasa@kek.jp\\

\vspace{5mm}

\end{center}
\vskip20pt
\begin{quotation}
\noindent
\begin{center}

{\bf Abstract}
\end{center}
We investigate a single scalar neutrino production at 
the upgraded HERA with high luminosity in the 
framework of an R-parity violating supersymmetric model. 
We find that 
the scalar neutrino with mass around $100$GeV could be observed 
through investigating the multilepton production process 
$e^- p \to \mu^-  \sneu_{\tau} X \to \mu^- (e^- \mu^+) X$. 
The signal would be characterized by $\mu^+$ and $e^-$ 
with high transverse momentum 
as well as a sharp peak in the invariant mass $M(e^-\mu^+)$ 
distribution. 
\end{quotation}
\vfill\eject
\section{\it Introduction}
\renewcommand{\thefootnote}{\fnsymbol{footnote}}
The $ep$ collider HERA is a unique accelerator because it 
enables us to investigate 
the proton structure in extremely deep region. 
Furthermore we can study 
the basic electron-quark interactions at the high energies only by 
HERA which provides us both baryonic and leptonic quantum numbers 
in the initial states. 
In fact HERA could set the best limit on the masses 
and the coupling strengths of leptoquarks via detailed analyses of 
the deep inelastic (DIS) neutral current (NC) and charged current 
(CC) processes. 

It is widely known that the scalar partners of quarks 
(squarks : $\squark$) have similar properties to the leptoquaks 
in a class of the supersymmetric (SUSY) standard models, the R-parity 
breaking (RB) models \cite{rbphys}. 
Here the R-parity is defined by $R = (-1)^{L + 3 B + 2 S}$, where 
$L$, $B$ and $S$ denote the lepton number, the baryon number and the spin, 
respectively. 
The models are characterized by the R-parity violating 
${\hat{L}}{\hat{Q}}{\hat{D}}$ type superpotentials, from which the 
leptoquark interactions $\ell -q -\squark$ are derived. 
The superfields ${\hat{L}}$, ${\hat{Q}}$ and ${\hat{D}}$ contain, 
respectively, the left-handed leptons, the left-handed quarks and 
the right-handed down-quark together with the SUSY partners 
$\sll$, $\sql$ and $\sdr$. 
We could study in detail the single production of squarks 
as well as leptoquarks by HERA experiments. 

  In 1997 both H1 \cite{H1} and ZEUS \cite{ZEUS} collaborations 
  reported an event excess at large $x$ and high $Q^2$ 
in comparison with the Standard Model (SM) expectations in the 
DIS $e^+p \to e^+ X$. 
The news very much excited the high energy physics community. 
Various ideas to understand the anomaly 
have extensively been examined by theoreticians since then 
\cite{rbphys,hewett,dreiner}. 
We have also proposed an interpretation of the anomalous events by the 
scalar top quark 
(stop) production in the SUSY models with RB interactions \cite{stoprb,dbst}. 
Contrary to initial expectation, the anomalies have gradually faded away from 
the whole data sample with increasing experimental data. 
However, this fact impressed us that HERA could have the potentiality exploring 
physics beyond the SM. 

In this paper we consider another type of the R-parity violating SUSY model, 
with the ${\hat{L}}{\hat{L}}{\hat{E}}$ type superpotential, 
where ${\hat{E}}$ contains 
the right-handed leptons together with the SUSY partners $\slr$. 
We show that the multilepton events come from 
the single production of a scalar partner of the neutrinos (sneutrinos) 
at HERA. 
As the process is not an $eq$ scattering of the resonant type, 
the number of events is too small  to be detected at the present HERA 
with the integrated luminosity of around $100$pb$^{-1}$. 
Fortunately, however, HERA has been upgraded,
and experiments using polarized  $e^-$ or $e^+$  beams with  high luminosity
will soon start\cite{schedule}.  
Then it will be expected that 
we are able to search for rare events with higher 
statistics than before.

\section{\it Models and constraints}
In the minimal SUSY standard model (MSSM), 
the general RB superpotential $W_{\rb}$ is written by  \cite{Barger}
\begin{equation}
W_{\rb}=\lambda_{ijk}\hat{L}_i \hat{L}_j \hat{E^c}_k 
+ \lambda'_{ijk}\hat{L}_i \hat{Q}_j \hat{D^c}_k + 
\lambda''_{ijk}\hat{U^c}_i \hat{D^c}_j \hat{D^c}_k, 
\label{RBW}
\end{equation}
where $i$, $j$ and $k$ are generation indices. 
The first two terms violate the lepton number $L$ and the last term 
violates the baryon number $B$. 
Incorporating RB interactions into the MSSM 
we have a possibility to unveil yet unresolved problems as 
({\romannumeral 1}) the cosmic baryon number violation, 
({\romannumeral 2}) the origin of the masses and the 
magnetic moments of neutrinos and 
({\romannumeral 3}) some interesting rare processes 
induced by the $L$ and/or $B$ violation. 
Here we consider the first ${\hat{L}_i}{\hat{L}_j}{\hat{E}_k}$ 
term, in which we set 
$(i, j, k) = (2,3,1)$. 
The Yukawa-type interaction Lagrangian
\begin{equation}
L=\lambda_{231} ( 
  {\overline{\widetilde{e_{R}}}}{\overline{\nu_\mu}} \tau_L 
-{\overline{\widetilde{e_{R}}}}{\overline{\nu_\tau}} \mu_L 
- {\widetilde{\nu_{\tau}}} \overline{e^c} \mu_L + \cdots) + h.c. 
\label{intrb}
\end{equation}
is derived from the superpotential from (1). 

The most stringent upper bound on $\lambda_{231}$ comes  
from the leptonic decay width of the tau lepton \cite{Barger}, 
\begin{equation}
\lambda_{231} \nle 0.07{\frac{m_{\widetilde{e_R}}}{100 GeV}}, 
\end{equation}
since the selectron exchange diagram through the first and second terms 
in (\ref{intrb}) 
contributes to the process $\tau \to \nu_\mu \nu_\tau \mu$. 
The upper limit is not so severe if we compare above bound (3) with 
the bounds for other RB couplings, e.g., 
$\lambda_{133}$ $\nle$ $0.0060$ $\sqrt{m_{\widetilde{\tau}}/100 {\rm GeV}}$ 
derived from the experimental limit on the electron neutrino mass
\cite{dreiner,godbole}. 
This is the reason why we consider the ${\hat{L}_2}{\hat{L}_3}{\hat{E}_1}$ 
superpotential.

We are aware of constraints on the slepton masses from the precision measurements 
at LEP2 \cite{leplimit}. 
The present lower mass limits at $95\%$ C.L. are
$m_{\sneu_{\mu,\tau}} \nge 65$GeV and $\mser \nge 69$GeV, which are obtained 
by the analyses on the pair production of the sleptons with 
the RB decay modes. 
In the following analysis we take typical input parameters 
$\msn=100$GeV, 
$\lambda_{231}=0.1$  
and additionally we assume 
$Br(\sneu_\tau \to e \mu)=1$ for simplicity. 
This assumption corresponds to the case of a large SU(2) gaugino mass $M_2$ and a 
large SUSY Higgs mass $\mu$, e.g., $M_2 > 300$GeV and $\mu > 100$GeV.

\section{\it Production processes}

We consider the single sneutrino production as a signal process, 
\begin{equation}
e^- p \to \mu^-  \sneu_{\tau} X. 
\label{sigprc}
\end{equation}
Note that the mass threshold of the process should be 
$\sqrt{s_{eq}}\nge \msn$ and it is lower than 
production threshold at $e^+e^-$ or $pp$ colliders as far as 
the $\sneu$ pair production is concerned. 
The $\sneu_\tau$ decays into $e^- \mu^+$
 via the R-parity violating interaction (\ref{intrb}). 
 Then we have the final state \footnote
   {The same final state can be expected when we take the non-zero 
   $\lambda_{121}$ or $\lambda_{122}$ couplings. 
   For this case $\sneu_e$ or $\sneu_\mu$ is singly produced with $\mu^-$ in the 
   $eq$ collisions. However, the magnitudes of $\lambda_{121}$ and $\lambda_{122}$
   are more severely constrained from the charged current universality than 
   $\lambda_{231}$ \cite{dreiner}}
\begin{equation}
e^- p \to \mu^- \mu^+ e^- X. 
\label{bgprc}
\end{equation}
Throughout the present work 
whole calculations of decay widths and cross sections 
have been  performed by using the {\tt GRACE/SUSY} system, 
an automatic computation 
program for SUSY processes \cite{sgrace}. 
While the {\tt GRACE/SUSY} system is originally designed to treat such 
elementary 
subprocesses as $e^+e^-$, $eq$ and $qq$ collisions, 
we have recently succeeded in extending it to $ep$ and $pp$ collisions. 
We use the extended new versions which 
include an interface to the PDFLIB too. 
For the parton distribution function we have used CTEQ4M \cite{cteq}. 
The Feynman diagrams of the subprocesses for(\ref{sigprc}) as well as 
relevant SM background are shown in Figs.1 and 2, respectively. 
\begin{figure}[hbtp]
\begin{center}
\epsfig{figure=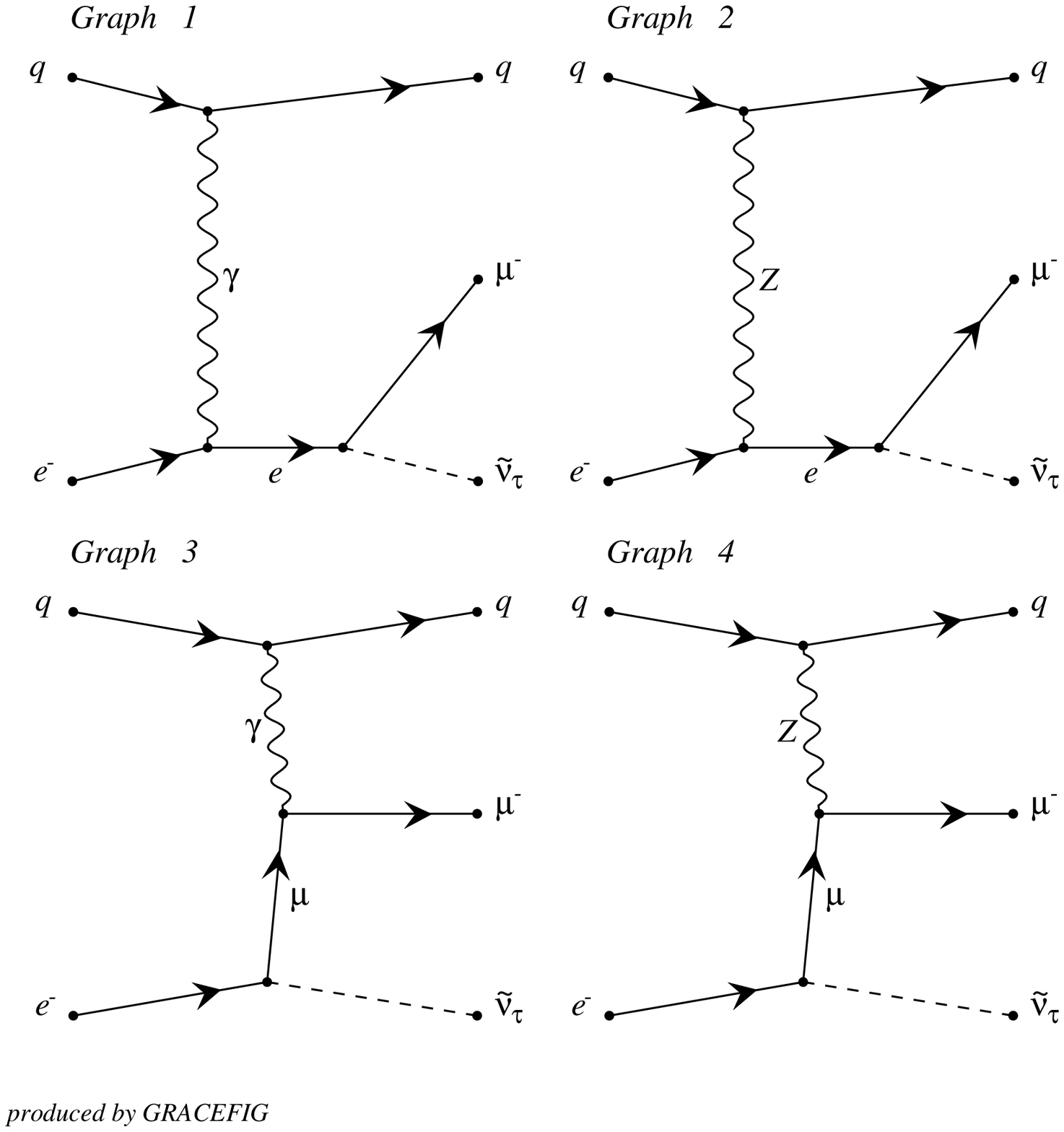,height=8cm,angle=0}
 \caption{
  Feynman diagrams for $e^- q \to \mu^-  \sneu_{\tau} q$ in the RB model.
  }
 \label{fig:rbfeynman}
\end{center}
\end{figure}
\begin{figure}[hbtp]
\begin{center}
\epsfig{figure=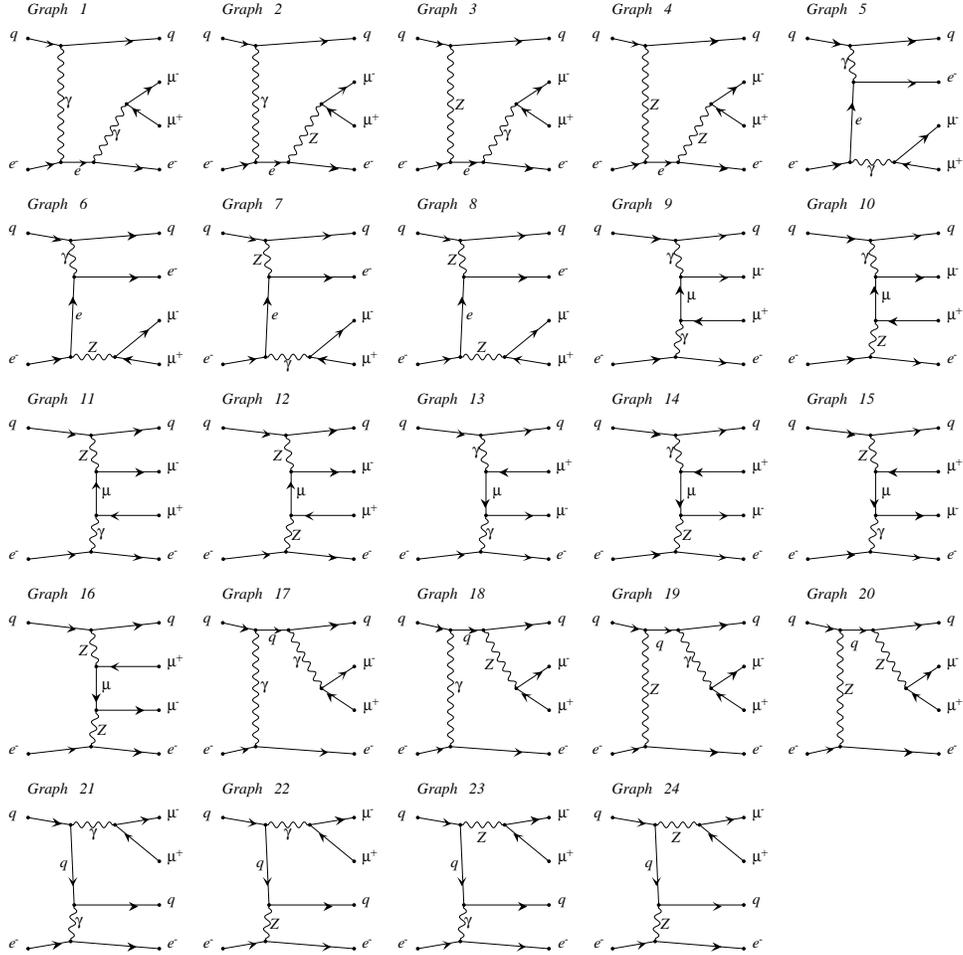,height=15cm,angle=0}
 \caption{
  Feynman diagrams for $e^- q \to \mu^- \mu^+ e^- q$ in the SM.
  }
 \label{fig:smfeynman}
\end{center}
\end{figure}

\section{\it Numerical results}
We obtain the total cross section of 10fb for the signal process 
(\ref{sigprc}), 
where we take 
$\lambda_{231}=0.1$,  $m_{\sneu_\tau}=100$GeV. 
It is too small cross section for us to extract from 
the SM background with the same 
final states $\sigma_{SM} (e^- p \to \mu^- \mu^+ e^- X) \sim 0.3$nb. 
Consequently we should suppress the huge background 
to extract the signal. 

In Fig.3 we show the various transverse momentum ($P_T(e^-)$,  
$P_T(\mu^-)$ and $P_T(\mu^+)$) distributions. 
We find that the signal cross section is 
much smaller than the SM background in the $P_T(\mu^-)$ distribution 
in a whole kinematical region. 
The main contribution to the SM background comes from 
the two photon processes (graphs 9, 13 in Fig.2)and 
the virtual photon emission processes (graphs 1, 5, 17, 21 in Fig.2). 
This is reflected in the rapid increase in the $P_T$ distribution with decreasing 
$P_T$ as shown in Fig.3. 
On the other hand, in the $P_T(e^-)$, $P_T(\mu^+)$ distributions, the signal 
differential cross section can be comparable to the background in a 
specific kinematical region, $P_T(e^-)$, $P_T(\mu^+)$ $\simeq$ $50$GeV. 
This is due to the fact that the $e^-$ and $\mu^+$ are originated from 
the 2-body decay 
of the sneutrino with mass $100$GeV. 
We can see the similar small Jacobian peak from the $Z$-boson decay 
for the SM histgrams in the 
$P_T(\mu^+)$ and $P_T(\mu^-)$ ($\simeq 46$GeV) distributions. 

\begin{figure}[hbt]
\begin{center}
\epsfig{figure=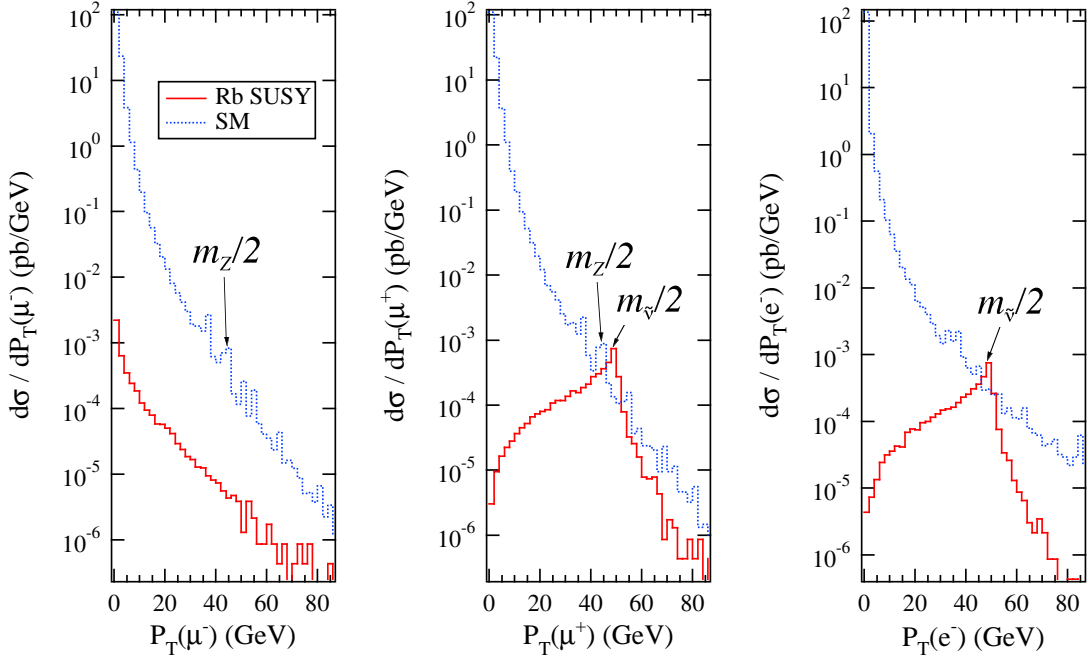,height=9cm,angle=0}
 \caption{
  Transverse momentum distributions for 
  $e^- p \to \mu^- \mu^+ e^- X$. 
  We take $\lambda_{231}$ $=$ $0.1$,  $m_{\sneu_\tau}$ $=$ $100$GeV. 
  }
 \label{fig:ptdist}
\end{center}
\end{figure}

The Monte Carlo event simulation 
in $(P_T(\mu^+),P_T(e^-))$ plane is displayed in Fig.4 
assuming the integrated luminosity of $1$fb$^{-1}$.
While the number of signal events is small compared to the backgrounds, 
they are characterized by the large transverse momentum. 
Based on the above observation, we expect that the background 
can be suppressed if we impose appropriate kinematical cuts on the 
final leptons. 
Specifically, we find following condition is suitable for the purpose, 
\begin{eqnarray}
P_T(\mu^+)>20{\rm GeV}, \\
P_T(e^-)>20{\rm GeV} .
\end{eqnarray}

\begin{figure}[hbt]
\begin{center}
\epsfig{figure=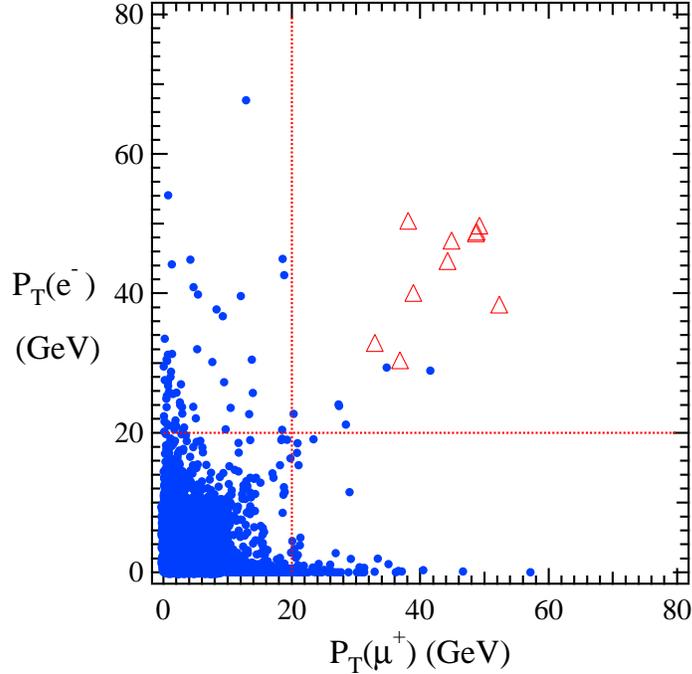,height=9cm,angle=0}
 \caption{
  Scatter plots of $P_T$ for 
  $e^- p \to \mu^- \mu^+ e^- X$. 
  Open triangle and closed circle respectively correspond to 
  $e^- p \to \mu^-  \sneu_{\tau} X \to \mu^- (e^- \mu^+) X$ in 
  the RB model and 
  $e^- p \to \mu^- \mu^+ e^- X$ in the SM. 
  We take $\lambda_{231}=0.1$,  $m_{\sneu_\tau}=100$GeV and 
  assume $L=1$fb$^{-1}$.
  }
 \label{fig:scatter}
\end{center}
\end{figure}

When we apply the $P_T$ cuts, we find that the signal can be seen as an event 
excess in the 
relevant invariant mass distributions. 
In Fig.5 we show 
$M(\mu^+\mu^-)$ and $M(e^-\mu^+)$ distributions with the $P_T$ cuts (6) and (7). 
The almost uniform excess for the $M(\mu^+\mu^-)$ and 
a sharp peak at $M(e^-\mu^+)=m_{\sneu_\tau}=100$GeV can be expected.
\begin{figure}[hbtp]
\begin{center}
\epsfig{figure=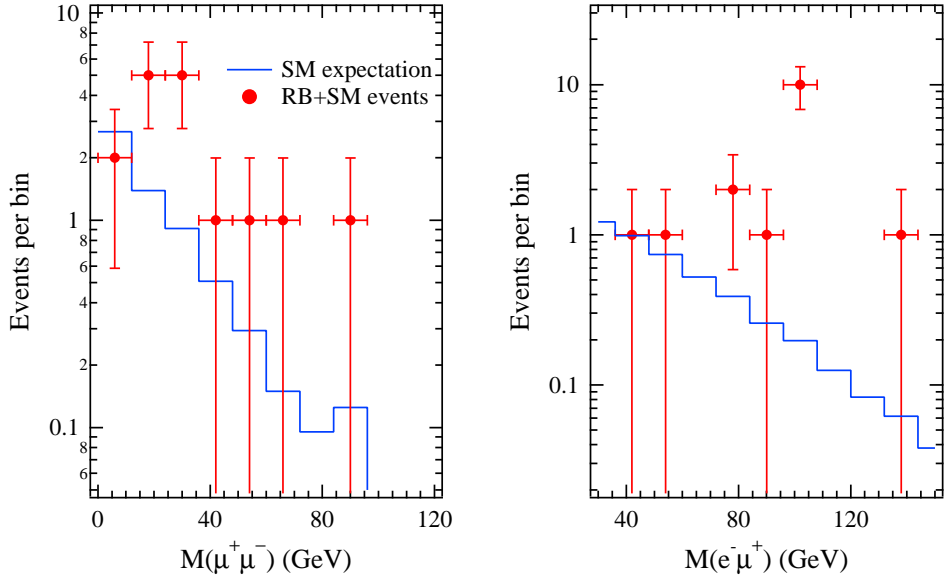,height=8cm,angle=0}
 \caption{
  Invariant mass distributions for 
  $e^- p \to \mu^- \mu^+ e^- X$ with cuts 
  $P_T(\mu^+)$, $P_T(e^-)$ $>$ $20{\rm GeV}$. 
  We take $\lambda_{231}=0.1$,  $m_{\sneu_\tau}=100$GeV and 
  assume $L=1$fb$^{-1}$.
  }
 \label{fig:invmass}
\end{center}
\end{figure}

\section{\it Summary and conclusion}
We have investigated the single scalar neutrino production at HERA in the 
framework of the R-parity breaking SUSY model with $\lambda_{231} \neq 0$.
The signal of the process should be multilepton final states 
$e^-\mu^+\mu^-$ without missing energies. 

The main background would be the basic QED reactions, i.e., 
two photon processes and the virtual photon emission from 
initial and final fermions.
We find that the background can be suppressed by making 
the appropriate $P_T$ cut for for final $e^-$ and $\mu^+$. 
Then the signal can be clearly seen as a sharp peak at $M(e^-\mu^+)=m_{\sneu_\tau}$
in the invariant mass distribution.

We conclude that if the R-parity violating coupling constant is 
$\lambda_{231}\nge 0.1$ and the mass of scalar neutrino is $m_{\sneu_\tau}\nge 100$GeV,
 the HERA could 
be efficient for exploring the ${\hat{L}}{\hat{L}}{\hat{E}}$ type 
interactions in the R-parity breaking SUSY model.

\vskip30pt
\begin{large}
{\bf Acknowledgements}
\end{large}

\noindent
The work  was supported in part by the Grant-in-Aid for Scientific Research
from the Ministry of Education, Culture, Sports, Science and Technology of Japan, 
No. 10440080 and No. 11206203.


\end{document}